\documentclass[aps,showpacs, prl,twocolumn,superscriptaddress]{revtex4}%
\usepackage{epsfig,dsfont,amssymb,amsmath,amsthm,amsfonts,amsbsy,mathrsfs}
\usepackage{graphicx}
\usepackage{amsmath}
\usepackage{amssymb}
\begin{document}
\title{New Superhard Carbon Phases Between Graphite and Diamond}
\author{Chaoyu He}
\affiliation{Institute for Quantum Engineering and Micro-Nano Energy
Technology, Xiangtan University, Xiangtan 411105, China}
\author{L. Z. Sun}
\email{lzsun@xtu.edu.cn} \affiliation{Institute for Quantum
Engineering and Micro-Nano Energy Technology, Xiangtan University,
Xiangtan 411105, China}
\author{C. X. Zhang}
\affiliation{Institute for Quantum Engineering and Micro-Nano Energy
Technology, Xiangtan University, Xiangtan 411105, China}
\author{K. W. Zhang}
\affiliation{Institute for Quantum Engineering and Micro-Nano Energy
Technology, Xiangtan University, Xiangtan 411105, China}
\author{Xiangyang Peng}
\affiliation{Institute for Quantum Engineering and Micro-Nano Energy
Technology, Xiangtan University, Xiangtan 411105, China}
\author{Jianxin Zhong}
\email{zhong.xtu@gmail.com}\affiliation{Institute for Quantum
Engineering and Micro-Nano Energy Technology, Xiangtan University,
Xiangtan 411105, China}
\date{\today}

\begin{abstract}
Two new carbon allotropes (H-carbon and S-carbon) are proposed, as
possible candidates for the intermediate superhard phases between
graphite and diamond obtained in the process of cold compressing
graphite, based on the results of first-principles calculations.
Both H-carbon and S-carbon are more stable than previously proposed
M-carbon and W-carbon and their bulk modulus are comparable to that
of diamond. H-carbon is an indirect-band-gap semiconductor with a
gap of 4.459 eV and S-carbon is a direct-band-gap semiconductor with
a gap of 4.343 eV. The transition pressure from cold compressing
graphite is 10.08 GPa and 5.93 Gpa for H-carbon and S-carbon,
respectively, which is in consistent with the recent experimental report.\\
\end{abstract}
\maketitle
\section{Introduction}
\indent Carbon is considered as the most active element in the
periodic table due to its broad sp, sp$^{2}$ and sp$^{3}$
hybridizing ability. Besides the four best-known carbon allotropes,
graphite, cubic-diamond, hexagonal diamond and amorphous carbon, an
unknown superhard phase of carbon has been reported in experiment
\cite{1, 2, 3, 4, 5} along with the structural phase transition in
cold compressed graphite. Several structures have been proposed as
the candidates for this superhard phase, such as the monoclinic
M-carbon \cite{61, 6}, cubic body center C4 carbon (bct-C4) \cite{7}
and the orthorhombic W-carbon \cite{8}. The cohesive energy of the
most stable W-carbon is about 160 meV per atom higher than that of
diamond, indicating that other more stable carbon phases may exist.
Very recently, another new carbon allotrope, named as Z-carbon, was
proposed and investigated at almost the same time by three
independent research groups \cite{9, 10, 11}. Z-carbon is more
stable (its cohesive energy is about 129 meV per atom larger than
that of diamond) and harder than W-carbon. Moreover, its transition
pressure is around 10 Gpa which is lower than that of W-carbon.
Thus, it is believed that the formation in the process of the cold
compressing graphite prefers Z-carbon rather than W-carbon. Although
none of them can fit the experimental results satisfactorily, these
theoretically proposed phases are significant in understanding
the experimental process of cold compressing of graphite.\\
\begin{figure}
\center
\includegraphics[width=2.8in]{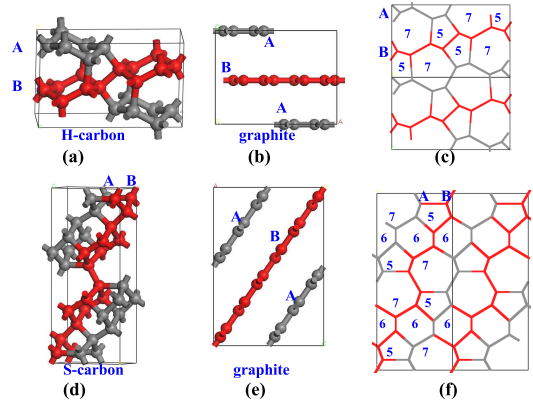}\\
\caption{Crystal structure of H-carbon(a), initial AB stacking
graphite supercell for H-carbon (b) and side view containing five
and seven carbon rings of H-carbon (c). Crystal structure of
S-carbon (d), initial AB stacking graphite supercell for S-carbon
(e) and side view containing five and seven carbon rings of S-carbon
(f).}\label{fig1}
\end{figure}
\indent The formation of the new carbon phase needs large lattice
distortions, such as buckling between adjacent carbon layers of
graphite. Therefore, the phase transition depends not only on the
particular nucleation history \cite{12} but also on the structure of
the starting graphitic material. Intuitively, such buckling between
carbon layers will start from the natural AB stacking graphite
rather than the AA stacking graphite chosen in previous theoretical
studies \cite{6, 7, 8, 9, 10, 11}. From this hypothesis, we propose
two new superhard crystals, namely H-carbon and S-carbon, as
potential candidates for the experimentally observed intermediate
state of carbon, which can be formed from compressing AB stacking
graphite. Our first-principles calculations based on density
functional theory show that both the allotropes are transparent wide
band gap insulators and their hardness are comparable to that of
diamond. S-carbon has the lowest enthalpy compared with bct-C4
carbon, M-carbon, W-carbon, and Z-carbon and is the most stable new
carbon phase theoretically proposed until now. H-carbon is always
more stable than W-carbon and M-carbon.
When the pressure is above 5.93 Gpa (10.08 Gpa), S-carbon (H-carbon) is more favorable than graphite.\\
\section{Models and Methods}
\indent H-carbon and S-carbon belong to the orthorhombic lattice and
they can be obtained from cold compressing AB stacking graphite.
Crystal structures of H-carbon and S-carbon are shown in
Fig.~\ref{fig1} (a) and (d), and their corresponding initial
graphite structures are shown in Fig.~\ref{fig1} (b) and (e),
respectively. H-carbon with the Pbam symmetry containing 16 carbon
atoms per crystal cell is obtained by reorganizing C atoms to form
sp$^3$ bonds through compressing an AB stacking graphite supercell
with 16 atoms. While, S-carbon with Cmcm symmetry containing
24 carbon atoms per crystal cell is obtained from larger AB stacking graphite supercell with 24 atoms.\\
\indent All calculations are carried out using the density
functional theory within both local density approximation (LDA)
\cite{13, 14} and general gradient approximation (GGA) \cite{15} as
implemented in Vienna ab initio simulation package (VASP) \cite{16,
17}. In evaluating of the transition pressure from graphite to each
superhard phases, it is known that LDA is a better choice because
LDA can give reasonable interlayer distances, mechanical properties
of graphite sheets due to a delicate error cancelation between
exchange and correlation in comparison with that of semi-local
generalized gradient approximation (GGA). The interactions between
nucleus and the 2s$^{2}$2p$^{2}$ valence electrons of carbon are
described by the projector augmented wave (PAW) method \cite{18,
19}. A plane-wave basis with a cutoff energy of 500 eV is used to
expand the wave functions. The Brillouin Zone (BZ) sample meshes are
set to be dense enough to ensure the accuracy of our calculations
\cite{20}. Crystal lattices and atom positions of graphite, diamond,
M-carbon, W-carbon, Z-carbon, H-carbon and S-carbon are fully
optimized up to the residual force on every atom less than 0.005
eV/{\AA} through the
conjugate-gradient algorithm. Vibration properties are calculated by using the phonon package \cite{21} with the forces calculated from VASP.\\
\begin{figure}
\includegraphics[width=3in]{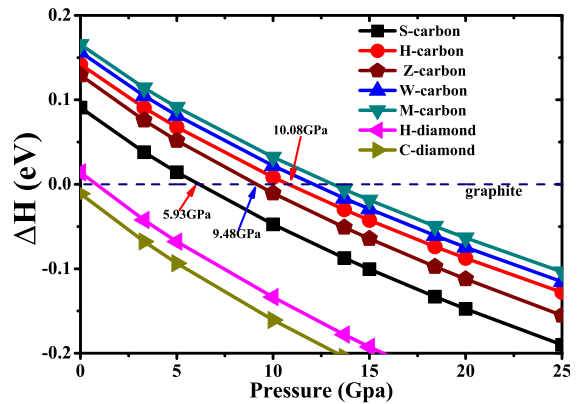}\\
\caption{The enthalpy per atom for cubic diamond, hexagonal diamond,
M-carbon, W-carbon, Z-carbon, H-carbon and S-carbon as a function of
pressure relative to graphite.}\label{fig2}
\end{figure}
\begin{table*}
  \centering
  \caption{Space group, exchange-correlation functional (EXC), lattice parameters (LP), density (g/cm$^{3}$), band gap (Eg: eV), cohesive energy (Ecoh: eV)
  and bulk modulus (B0: Gpa) for diamond, M-carbon, W-carbon, Z-carbon, H-carbon and S-carbon.}\label{tabI}
\begin{tabular}{c c c c c c c c}
\hline \hline
Systems &Space group &EXC &LP &Density &Eg  &Ecoh &B0\\
\hline
diamond &Fd-3m &LDA &a=b=c=3.536{\AA} &3.611 &4.648 &-8.997 &514.08\\
diamond &Fd-3m &GGA &a=b=c=3.574{\AA} &3.491 &4.635 &-7.693 &478.96\\
M-carbon &C2/m &LDA &a=9.093{\AA}, b=2.498{\AA}, c=4.108{\AA}, beta=96.96$^{\circ}$ &3.443 &3.532 &-8.821 &447.33\\
M-carbon &C2/m &GGA &a=9.194{\AA}, b=2.525{\AA}, c=4.151{\AA}, beta=97.03$^{\circ}$ &3.333 &3.493 &-7.531 &431.44\\
W-carbon &Pnma &LDA &a=8.984{\AA}, b=4.116{\AA}, c=2.498{\AA} &3.455 &4.325 &-8.830 &468.12\\
W-carbon &Pnma &GGA &a=9.086{\AA}, b=4.156{\AA}, c=2.525{\AA} &3.345 &4.281 &-7.539 &447.03\\
Z-carbon &Cmmm &LDA &a=8.677{\AA}, b=4.211{\AA}, c=2.489{\AA} &3.507 &3.414 &-8.857 &497.87\\
Z-carbon &Cmmm &GGA &a=8.772{\AA}, b=4.256{\AA}, c=2.514{\AA} &3.394 &3.273 &-7.564 &464.13\\
H-carbon &Pbam &LDA &a=7.792{\AA}, b=4.757{\AA}, c=2.497{\AA} &3.440 &4.512 &-8.845 &466.92\\
H-carbon &Pbam &GGA &a=7.874{\AA}, b=4.807{\AA}, c=2.524{\AA} &3.339 &4.459 &-7.554 &445.88\\
S-carbon &Cmcm &LDA &a=2.496{\AA}, b=11.293{\AA},c=4.857{\AA} &3.489 &4.451 &-8.896 &486.29\\
S-carbon &Cmcm &GGA &a=2.523{\AA}, b=11.385{\AA},c=4.899{\AA} &3.399 &4.342 &-7.593 &468.45\\
\hline \hline
\end{tabular}
\end{table*}
\section{Results and Discussion}
\begin{figure}
\includegraphics[width=3.50in]{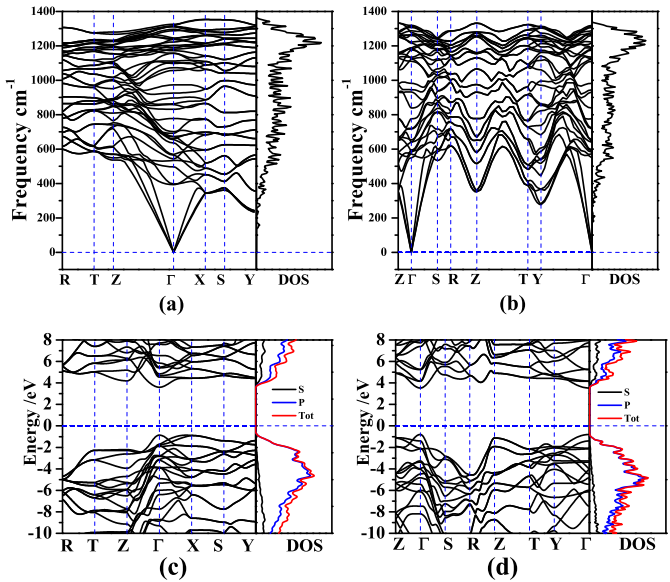}
\caption{Phonon band structure and density of state for H-carbon (a)
and S-carbon (b). Electronic band structure and density of state for
H-carbon (c) and S-carbon (d).}\label{fig3}
\end{figure}

\begin{figure}
\includegraphics[width=3.0in]{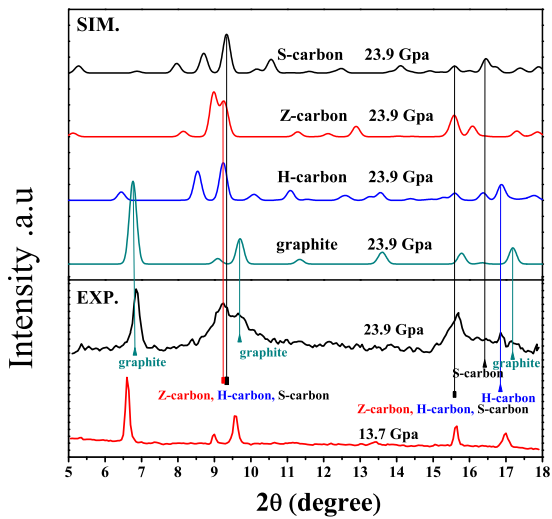}\\
\caption{Simulated XRD data for S-carbon, Z-carbon, H-carbon, and
graphite at corresponding pressure as well  as the experimental XRD
data at 23.9 Gpa and 13.7 GPa for the superhard intermediate state
derived from reference [5].}\label{fig4}
\end{figure}

\indent Similar to M-carbon and W-carbon, H-carbon and S-carbon
contain distorted five and seven carbon rings, as shown in
Fig.~\ref{fig1} (c) and (f). Their difference from Z-carbon and
bct-C4 carbon is the absence of four and eight carbon rings. The
crystal structure of H-carbon belongs to the Pbam space group. At
zero pressure, the GGA calculated equilibrium lattice constants are
a=7.874 {\AA}, b=4.807{\AA} and c=2.524 {\AA}. Four inequivalent
atoms in this crystal occupy the positions at (0.046, 0.361, 0.500),
(0.157, 0.309, 0.000), (0.327, 0.462, 0.000) and (0.430, 0.393,
0.500), respectively. S-carbon belongs to the Cmcm space group and
its equilibrium lattice constants from GGA calculations are a=2.523
{\AA}, b=11.385 {\AA} and c=4.899 {\AA}. Four inequivalent atoms in
S-carbon locate at positions of (0.500, 0.778, 0.250), (0.500,
0.442, 0.078), (0.500, 0.132, 0.518) and (0.000, 0.300, 0.750).
Recently proposed Z-carbon contains even number carbon rings and
hold Cmmm symmetry, with equilibrium lattice parameters a=8.772
{\AA}, b=4.256 {\AA} and c=2.514 {\AA}. There are only two
inequivalent atoms in Z-carbon located at (0.089, 0.316, 0.500) and
(0.167, 0.185, 0.000). All the results of the lattice constants of
diamond, M-carbon, W-carbon, Z-carbon, H-carbon and S-carbon derived
from both GGA and LDA calculations are listed in Tab.~\ref{tabI}.
The LDA calculations tend to give smaller lattice constants for all
structures considered here, but the conclusions derived from LDA and GGA are consistent.\\
\indent The relative stability of diamond, M-carbon, W-carbon,
Z-carbon, H-carbon and S-carbon is evaluated through comparing their
cohesive energy per atom. The energy of H-carbon is about 15 meV
lower than that of W-carbon, about 25 meV lower than that of
M-carbon, and 10 meV higher than that of Z-carbon at both LDA and
GGA level. Namely, H-carbon is more stable than M-carbon and
W-carbon but less stable than Z-carbon. Among these allotropes,
S-carbon is the most stable one whose cohesive energy is about 30
meV lower than that of Z-carbon. From the cohesive energy, S-carbon
is the most stable carbon allotrope theoretically proposed until
now. The enthalpy per atom for cubic diamond, hexagonal diamond,
M-carbon, W-carbon, Z-carbon, H-carbon as well as S-carbon as a
function of pressure relative to graphite derived from LDA
calculations are shown in Fig.~\ref{fig2}. The results indicate that
H-carbon is more stable than graphite when the external pressure is
larger than 10.08 GPa. H-carbon is always favorable than M-carbon
and W-carbon at the pressure range from 0 to 25 GPa. Amazingly, the
transition pressure point of S-carbon is only 5.93 Gpa, which is
smaller than all of the previously proposed carbon allotropes. To
further confirm the dynamic stabilities of H-carbon and S-carbon, we
calculate their phonon band structures and phonon densities of state
as shown in Fig.~\ref{fig3}. Both LDA and GGA calculations indicate
that there is no negative frequency for both structures confirming
that H-carbon and S-carbon
are dynamic stable phases of carbon.\\
\indent Space group, density, band gap, cohesive energy and bulk
modulus of diamond, M-carbon, W-carbon, Z-carbon, H-carbon, and
S-carbon are also summarized in Tab.~\ref{tabI}. The results
indicate that, as superhard intermediate phases between graphite and
diamond, M-carbon, W-carbon, Z-carbon, H-carbon, and S-carbon can be
formed by cold compressing graphite and their densities and bulk
modulus are close to that of diamond. The density of H-carbon and
S-carbon derived from GGA is 3.339 g/cm$^{3}$ and 3.399 g/cm$^{3}$,
respectively, which is similar to that of M-carbon (3.333
g/cm$^{3}$), W-carbon (3.345 g/cm$^{3}$) and Z-carbon (3.394
g/cm$^{3}$) derived from the same method. The bulk modulus of
H-carbon is 445.88 GPa which is almost the same as that of W-carbon
(447.03 GPa) and slightly larger than that of M-carbon (431.44 GPa).
The bulk modulus of S-carbon is 468.45 Gpa which is close to that of
Z-carbon (466.13 Gpa). The values of the
bulk modulus indicate that H-carbon and S-carbon are superhard materials which are comparable to diamond (478.96 Gpa). \\
\indent Electronic properties of diamond, M-carbon, W-carbon,
Z-carbon, H-carbon and S-carbon are also investigated under both GGA
and LDA calculations. All these superhard carbon allotropes are
indirect wide band gap semiconductors except for S-carbon. The GGA
results show that H-carbon holds an indirect wide gap of 4.531 eV
which is larger than that of M-carbon (3.523 eV ), W-carbon (4.325
eV ) and Z-carbon (3.273 eV), and smaller than that of diamond
(4.635 eV). Different from other previously proposed superhard
carbon allotropes, S-carbon is a direct-band-gap semiconductor with
a gap of 4.342 eV. The wide band gap indicates that both H-carbon
and S-carbon are transparent carbon allotropes. Fig. 3(c) and (d)
show the electronic band structures and densities of states of
H-carbon and S-carbon derived from GGA. The densities of states
indicate that the valence band maximum for both H-carbon and
S-carbon is mainly contributed from the
2p states of carbon, whereas the conduction band minimum comes from the sp hybridized states.\\
\indent The simulated X-ray diffraction (XRD) patterns for graphite,
Z-carbon, H-carbon and S-carbon at the pressure of 23.9 Gpa are
shown in Fig.~\ref{fig4} to compare with the experimental data from
reference [5]. The experimental data under high pressure can be
explained by H-carbon and S-carbon to some extent. The main XRD
peaks located in the region between $8.5^{o}$- $11^{o}$ and
$15^{o}$- $17^{o}$ for H-carbon and S-carbon agree well with the
experimental observations. As indicated by Amsler et al.\cite{10}
that sole comparison between experimental and theoretical XRD can
not directly clarify the new carbon phase. In view of the cohesive
energy, dynamical stability, electronic structure, bulk modulus,
lowest phase transition pressure point, and especially
the natural transition from AB stacking graphite, H-carbon and S-carbon are excellent candidates for the new superhard carbon phase. \\
\indent \textbf{Note. added} During the course of our submission, we
became aware of several recent papers \cite{n1, n2, n3, n4, n5, n6,
n7, n8} reporting the new superhard carbon phases similar to our H-carbon \cite{n1, n4} and S-carbon \cite{n2, n4}\\
\section{Conclusion}
\indent  In summary, we propose two new superhard carbon phases,
H-carbon and S-carbon, as  the possible candidates for the
intermediate phase of cold compressing graphite. Both H-carbon and
S-carbon with sp$^{3}$ carbon bonds are optical transparent
superhard carbon phases. They are more stable than M-carbon and
W-carbon, and S-carbon is the most table carbon phase theoretically
proposed until now. Moreover, when the pressure is above  5.93 Gpa
(10.08 Gpa), S-carbon (H-carbon) is more favorable than graphite.
These two new members together with the previous proposed M-carbon,
bct-C4, W-carbon and Z-carbon will enrich the theoretical evidence for understanding the experimental observation.\\
\section*{Acknowledgement}
This work is supported by the National Natural Science Foundation of
China (Grant Nos. 11074211, 10874143 and 10974166), the Cultivation
Fund of the Key Scientific and Technical Innovation Project, the Program for
 New Century Excellent Talents in University (Grant No. NCET-10-0169), and the
 Scientific Research Fund of Hunan Provincial Education Department (Grant Nos. 10K065, 10A118, 09K033) \\

\end{document}